\documentclass[11pt,epsf]{article}

\usepackage{epsfig}
\usepackage{amssymb}
\usepackage{graphicx}
\usepackage{color}
\usepackage{subfigure}

\begin{document}

\baselineskip 6mm
\renewcommand{\thefootnote}{\fnsymbol{footnote}}


\newcommand{\nc}{\newcommand}
\newcommand{\rnc}{\renewcommand}



\newcommand{\tcb}{\textcolor{blue}}
\newcommand{\tcr}{\textcolor{red}}
\newcommand{\tcg}{\textcolor{green}}


\def\be{\begin{equation}}
\def\ee{\end{equation}}
\def\ba{\begin{array}}
\def\ea{\end{array}}
\def\bea{\begin{eqnarray}}
\def\eea{\end{eqnarray}}
\def\nn{\nonumber\\}


\def\ct{\cite}
\def\la{\label}
\def\eq#1{(\ref{#1})}


\def\a{\alpha}
\def\b{\beta}
\def\g{\gamma}
\def\G{\Gamma}
\def\d{\delta}
\def\D{\Delta}
\def\ep{\epsilon}
\def\e{\eta}
\def\ph{\phi}
\def\Ph{\Phi}
\def\ps{\psi}
\def\Ps{\Psi}
\def\k{\kappa}
\def\l{\lambda}
\def\L{\Lambda}
\def\m{\mu}
\def\n{\nu}
\def\th{\theta}
\def\Th{\Theta}
\def\r{\rho}
\def\s{\sigma}
\def\S{\Sigma}
\def\ta{\tau}
\def\o{\omega}
\def\O{\Omega}
\def\pr{\prime}


\def\half{\frac{1}{2}}

\def\goto{\rightarrow}

\def\na{\nabla}
\def\grad{\nabla}
\def\curl{\nabla\times}
\def\div{\nabla\cdot}
\def\pa{\partial}

\def\bra{\left\langle}
\def\ket{\right\rangle}
\def\lb{\left[}
\def\lc{\left\{}
\def\ls{\left(}
\def\lp{\left.}
\def\rp{\right.}
\def\rb{\right]}
\def\rc{\right\}}
\def\rs{\right)}

\def\vac#1{\mid #1 \rangle}


\def\td#1{\tilde{#1}}
\def\check{ \maltese {\bf Check!}}


\def\Tr{{\rm Tr}\,}
\def\det{{\rm det}}


\def\bc#1{\nnindent {\bf $\bullet$ #1} \\ }
\def\ch {$<Check!>$ }
\def\ss {\vspace{1.5cm}}

\begin{titlepage}

\hfill\parbox{5cm} { }

\vspace{25mm}

\begin{center}
{\Large \bf  Correlation function of dyonic strings}

\vskip 1. cm
  { Xiaojian Bai$^{a}$ \footnote{e-mail : baixj87@sogang.ac.kr}
  Bum-Hoon Lee$^{ab}$\footnote{e-mail : bhl@sogang.ac.kr}
  and Chanyong Park$^b$\footnote{e-mail : cyong21@sogang.ac.kr}
  }

\vskip 0.5cm

{\it $^a\,$Department of Physics,, Sogang University, Seoul 121-742, Korea}\\
{ \it $^b\,$Center for Quantum Spacetime (CQUeST), Sogang University, Seoul 121-742, Korea}\\

\end{center}

\thispagestyle{empty}

\vskip2cm


\centerline{\bf ABSTRACT} \vskip 4mm

\vspace{1cm}
We investigate the two- and three-point correlation functions of the dyonic magnon and spike,
which correspond to the solitonic string moving in the Poincare $AdS$ and three-dimensional
sphere. We show that the coupling between two dyonic magnons or spikes together with a marginal
scalar operator in the string theory is exactly the same as one obtained by the RG
analysis in the gauge theory.

\vspace{2cm}


\end{titlepage}

\renewcommand{\thefootnote}{\arabic{footnote}}
\setcounter{footnote}{0}

\tableofcontents

\section{Introduction}

Recently, there was a big progress in calculating the three-point correlation function of heavy
operators like the magnon and spike with a marginal scalar operator. The information of this
three-point correlation function together with the conformal dimensions of all primary operators
is very important to characterize the conformal field theory (CFT). One of the well-known examples
for interacting CFT is the ${\cal N}=4$ super Yang-Mills (SYM) theory in four-dimensional space.
According to the AdS/CFT correspondence \ct{mal1} there exists a dual gravity theory having one-to-one
correspondence to the ${\cal N}=4$ SYM theory. This AdS/CFT correspondence provides new fascinating
way to understand the strong coupling region of SYM theory by investigating the dual classical gravity
or string theory. By many authors, the three-point correlation function of various heavy operators
were calculated and showed that the string theory calculations for the coupling between two heavy
operators and one marginal operator exactly coincide with the results of the gauge theory
\cite{Janik:2010gc}-\cite{Georgiou:2010an}. In this
paper, our goal is to investigate the correlation functions of more non-trivial solitonic strings
moving in $AdS_5 \times S^3$ and to compare those results with ones of the gauge theory.

After the integrable structure of the ${\cal N}=4$ SYM was founded \cite{Minahan:2002ve,Beisert:2003xu,Beisert:2003yb}, there were many works
in which the conformal dimensions of the various heavy operators of the gauge theory were studied
by investigating the dual solitonic string configurations by using the AdS/CFT correspondence
\cite{Ishizeki:2007we}-\cite{Gromov:2009bc}.
Also, such works were extended to the three-dimensional Chern-Simons theory, so called ABJM model
\cite{Aharony:2008ug}-\cite{Abbott:2009um}.
In these calculations, the magnon solution in the gauge theory was described by a solitonic
string solution moving in $R \times S^2$, where $R$ is the time direction in the global $AdS_5$ and
$S^2$ is a subspace of $S^5$.
This dual string configuration corresponding to the magnon was usually called the giant magnon.
Since the rank of the isometry group for $S^2$ is one, the dispersion relation of the giant magnon
is described by two conserved quantities, the energy and the angular momentum
\be
E - J = \frac{\sqrt{\l}}{\pi} \sin \frac{p}{2} .
\ee
This result corresponds to the magnon's dispersion relation of the spin chain model in the
large 't Hooft coupling limit. Furthermore, it was shown that there exists another solitonic string
solution, so called spike, whose dispersion relations was also investigated.

This giant magnon solution can be generalized to more complicated one like the solitonic string
moving in $R \times S^3$, which was called the dyonic magnon solution. Since the rank of the
isometry group for $S^3$ is two, the dispersion relation of the dyonic magnon can be described by
three conserved charges, one energy and two angular momenta \cite{Dorey:2006dq,Chen:2006gea}
\be
E - J_1 = \sqrt{J_2^2 + \frac{\l}{\pi^2} \sin^2 \frac{p}{2}} .
\ee
This is exactly the same as the dispersion relation of the magnon's bound state in the spin chain
model if we identify the second angular momentum $J_2$ with the number of magnons. Especially,
if we set $J_2 = 1$, it gives the magnon's dispersion relation calculated in the spin chain model.
Also, we can find the dyonic spike solution by choosing the different parameter region.

Recently, after regarding
the string moving in the Poincare $AdS_5 \times S^5$, the dispersion relation of various string
solutions like the spinning string, magnon and spike were investigated in the semi-classical limit 
\cite{Janik:2010gc}-\cite{Georgiou:2010an}.
Furthermore, the three-point correlation function between two heavy operators and one marginal
operator was also calculated, where heavy operator implies the magnon or spike.
Interestingly, the
coupling between these operators, which corresponds to the structure constant in the three-point
correlation function, was also calculated by the RG analysis in the gauge theory side. In the
previous works, it was shown that the string theory calculation of this coupling is perfectly
matched with the result of the RG analysis.
In this paper, we will investigate the correlation functions of the more general solitonic string
like the dyonic magnon or spike in the string theory and check the consistency of
the string calculation by comparing with the gauge theory result.

The rest part is organized as follows. In Sec. 2, we will consider a dyonic string moving in
$AdS_5 \times S^3$ and derive the equations of motion.
Using these equations of motion, we will calculate the two- and three-point correlation functions
for the dyonic magnon in Sec. 3 and the dyonic spike in Sec. 4. We will also check that the
structure constant of the three-point correlation function coincide with the gauge theory
results obtained by the RG analysis. Finally, we will finish our works with a brief discussion
in Sec.5.

{\bf Note added} At the final stage of this work, we noticed that there were some overlaps in 
Ref. \cite{Hernandez:2011up} for calculating the three-point correlation function of the dyonic magnon .

\section{Dyonic string on $AdS_5 \times S^3$}

Consider a solitonic string moving in $AdS_5 \times S^3$, which is a subspace of
$AdS_5 \times S^5$. In the global patch, while the dyonic string rotates on $S^3$
it is located at the center of $AdS$ space, which is dual to the dyonic magnon or spike
of the dual spin chain model. In the Poincare patch, the dyonic string moves as a point
particle in AdS bulk.
In the $AdS_5 \times S^3$ space having the following Euclidean metric
\be
ds^2 = \frac{R^2}{z^2} \ls dz^2 + d x^2 \rs + R^2 \ls d \th^2 + \sin^2 \th \ d \ph_1^2
+ \cos^2 \th  \ d \ph_2^2 \rs ,
\ee
the action describing the motion of the dyonic string becomes
\bea \label{E:stringaction}
	S_{st} &=& \frac{T}{2} \int d^2\sigma  \left( \frac{1}{z^2} \lc ( \pa_{\ta} {x})^2
+ ( \pa_{\ta} {z} )^2 \rc +  ( \pa_{\ta} {\theta})^{2}-   ( \pa_{\s} {\theta})^{2}
       + \sin^2\theta \lc {( \pa_{\ta}{\phi_1}})^2-   ( \pa_{\ta} {\phi_1})^2) \rc \rp \nn
       && \lp \qquad \qquad + \cos^2 \theta \lc {( \pa_{\ta}{\phi_2}})^2
       -   ( \pa_{\ta} {\phi_2})^2  \rc \frac{}{} \right),
\eea
where the string tension $T$ is related to the t' Hooft coupling like
\be	\la{rel:tension}
T \equiv 2 g =  \frac{\sqrt{\lambda}}{2 \pi} .
\ee
In the above, since the string moves as a point particle in the $AdS$ space, we take
$z$ and $x$ as a function of $\ta$ only.

The solutions describing the motion of the string in the $AdS$ part are given by
\bea
	z(\tau) &=& \frac{R}{\cosh \kappa\tau}  \nn
	x(\tau) &=& R\tanh \kappa \tau + x_0 ,
\eea
which are the specific parametrization of a geodesic $ \left(x(\tau)-x_0\right)^2
+ z(\tau)^2 = R^2 $ in $ AdS $. After substituting the above solution into the $AdS$
part action we obtain
\be\label{E:sads}
		S_{AdS}=\frac{T}{2} \int^{s/2}_{-s/2}d\tau\int^{L}_{-L}d\sigma\,\frac{1}{z^2}
(\dot{x}^2 + \dot{z}^2) = \kappa^2sLT.
\ee
By imposing the boundary conditions
\be
	\left(x(-s/2),z(-s/2)\right) = (0,\epsilon)\quad {\rm and}\quad \left(x(s/2),z(s/2)\right)
= (x_f,\epsilon)
\ee
where $ \epsilon $ is a very small number corresponding to the appropriate UV cut-off, we can
find
\be\label{E:kappa}
	\kappa \approx \frac{2}{s}\log\frac{x_f}{\epsilon}
\ee
with $ x_f \approx 2R \approx 2x_0 $.

Now, we consider the motion of the dyonic string rotating on $S^3$. To describe this
dyonic string, we choose the following ansatz
\be
	\theta = \theta(y),\quad \phi_1 = \nu_1\tau + g_1(y),\quad
	{\rm and} \quad \phi_2 = \nu_2\tau + g_2(y)
\ee
with
\be
y = a\tau + b\sigma .
\ee
Since there exist two rotation symmetries in the $\ph_1$- and $\ph_2$-directions,
the equations of motion for them are simply reduced to the total derivative
forms
\bea
	0 &=& \partial_y \lb \sin^2\theta \left( a\nu_1 + (a^2-b^2) {g}_1' \right) \rb , \nn
	0 &=& \partial_y \lb \cos^2\theta \left( a\nu_2 + (a^2-b^2) {g}_2' \right) \rb .
\eea
where the prime stands for the derivative with respect to $y$.
Then, $g_1'$ and $g_2'$ can be easily expressed by functions of $\theta$
\bea
	g_1' &=& \frac{1}{b^2-a^2}\left(a\nu_1 - \frac{c_1}{\sin^2\theta}\right)  ,\la{eq:ph1} \\
	g_2' &=& \frac{1}{b^2-a^2}\left(a\nu_2 - \frac{c_2}{\cos^2\theta}\right) , \la{eq:ph2}
\eea
where $ c_1 $ and $ c_2 $ are two integration constants.

The equation of motion for $ \theta $, after multiplying $ 2 \th'$, can be put into the following form
\be
	0 = \partial_y\left( \theta'^2 + \frac{1}{(b^2-a^2)^2}\left(b^2(\nu_1^2-\nu_2^2)\sin^2\theta
+ \frac{c_1^2}{\sin^2\theta}+ \frac{c_2^2}{\cos^2\theta}\right)\right) ,
\ee
from which we also reexpress $\th'$ in terms of $ \th $
\be\label{E:htheta}
	\th'^2 = -\frac{b^2(\nu_1^2-\nu_2^2)}{(b^2-a^2)^2\sin^2\theta}\left(\sin^4\theta
-  \frac{w^2\sin^2\theta}{b^2(\nu_1^2-\nu_2^2)} + \frac{c_1^2}{b^2(\nu_1^2-\nu_2^2)}
+ \frac{c_2^2 \tan^2\theta}{b^2(\nu_1^2-\nu_2^2)}\right),
\ee
where $\frac{w^2}{(b^2-a^2)^2}$ is introduced as another integration constant.

Notice that in the above there are three equations of motion for $\th$, $\ph_1$ and $\ph_2$,
so that we should generally determine six integration constants by imposing six boundary
conditions. Here, our aim is to find the dispersion relation of the dyonic string, which
is described by the three conserved quantities including one derivative. So, if we know
three integration constants $c_1$, $c_2$ and $w$, we can exactly determine the dispersion
relation of the dyonic string solution. As a result, only three boundary conditions are
required to determine the dispersion relation of the dyonic string and the other three
integration constants
are irrelevant.

\section{Dyonic magnon}

In this section, we will investigate the dyonic magnon solution, which corresponds to the
bound state of magnons in the dual gauge theory.
To find the dispersion relation of the dyonic magnon, we should first fix three integration
constants, $c_1$, $c_2$, and $w$.
To give the correct boundary conditions for dyonic magnon, we should know the typical
structure of the dyonic magnon's dispersion relation. Usually, in the infinite size limit
($\th_{max} = \frac{\pi}{2}$
or $L \to \infty$) the energy $E$ and the first angular momentum $J_1$
have the infinite values. On the other hand, the difference of them $E-J_1$, proportional to the string
world momentum $p$, and the second angular momentum $J_2$ are finite.
In the infinite size limit, we should  set $c_2=0$ to make the second angular momentum $J_2$
finite. If not, $g_2'$ in \eq{eq:ph2} diverges at $\th_{max}=\frac{\pi}{2}$.
The other boundary conditions for $\th$ and $\ph_1$ are the same as ones of the magnon
case \cite{Park:2010vs}, in which the solitonic string rotates on $S^2$. This implies that
when setting $\n_2 =0$
the dyonic magnon rotating on $S^3$ is reduced to the magnon on $S^2$.

For $\th$, we impose that there exists a maximum value $\th_{max}$ where $\th_{max}'=0$.
Then, \eq{E:htheta} can be rewritten as
\be	\la{eq:thmm}
\theta'^2 = \frac{b^2(\nu_1^2-\nu_2^2)}{(b^2-a^2)^2\sin^2\theta}\left(\sin^2\theta_{max}
-\sin^2\theta\right)\left( \sin^2\theta - \sin^2\theta_{min}\right),		
\ee
with
\bea
\sin^2 \th_{max} + \sin^2 \th_{min} &=& \frac{w^2}{b^2(\nu_1^2-\nu_2^2)}  , \nn
\sin^2 \th_{max} \cdot \sin^2 \th_{min} &=&  \frac{c_1^2}{b^2(\nu_1^2-\nu_2^2)} .
\eea
Imposing the boundary condition for $\ph_1$, $\pa_{\s} \ph_1 = 0$ at $\th_{max}$,
to \eq{eq:ph1}
gives rise to
\be	\la{eq:thmax}
\sin^2\theta_{max}= \frac{c_1}{a\nu_1} .
\ee
Substituting this result into \eq{E:htheta}, we can determine the integration constant $w^2$
as
\be
w^2=\frac{b^2c_1}{a\nu_1}(\nu_1^2-\nu_2^2) + a\nu_1c_1 ,
\ee
which means that $\sin^2 \th_{min}$ is given by
\be
\sin^2\theta_{min} = \frac{a \nu_1 c_1}{b^2(\nu_1^2-\nu_2^2)} .
\ee

From now on, we consider the infinite size limit only, where $\th_{max} = \frac{\pi}{2}$ and
$c_1 = a \n_1$ from \eq{eq:thmax}, so the minimum value of $\th$ in the this
limit is reduced to
\be
\sin^2\theta_{min} = \frac{a^2 \nu_1^2 }{b^2(\nu_1^2-\nu_2^2)} .
\ee
In the above, since $0<\sin^2\theta_{min} <1$, we can see that at least $\n_1 > \n_2$ and
$b > a$ where we assume that all parameters are positive.

\subsection{Two-point correlation function}

After convolution with the relevant wave function by following ref.\cite{janik}, the new action
in the infinite size limit $L \to \infty$ is given by
\bea\label{E:sbar}
	\bar{S} &= &S_{S^3} - \Pi_{\theta}\dot{\theta} - \Pi_{\phi_1}\phi_1 - \Pi_{\phi_2}\phi_2 \nn
			&=& -\frac{T}{2}\int^{s/2}_{-s/2}d\tau \int^{L}_{-L}d\sigma \, \n_1^2 \equiv -\rho^2sLT ,
\eea
where the energy of the solitonic string rotating on $S^3$ can be defined by the integration of $\r$
\bea\la{def:magnonenergy}
E &\equiv& T\int^{L}_{-L}d\sigma\,\rho \nn
	&=& 2T\frac{(b^2-a^2)\nu_1}{b^2\sqrt{\nu_1^2-\nu_2^2}}
	     \int^{\pi/2}_{\theta_{min}}d\theta\,\frac{\sin\theta}{\cos\theta}
\frac{1}{\sqrt{\sin^2\theta-\sin^2\theta_{min}}}.
\eea
In the second line in \eq{def:magnonenergy}, the energy is expressed by the integration
of the target space variable $\th$ and the infinite size limit $L \to \infty$ corresponds
to taking $\th_{max} \to \frac{\pi}{2}$, which means that the dyonic magnon in the dual
gauge theory has an infinite size.

Combining \eq{E:sads} and \eq{E:sbar} together with \eq{E:kappa}, the total action for the
dyonic magnon becomes
\be
	iS_{tot} \equiv i(S_{AdS} + \bar{S}) = i\left(\frac{4}{s^2}\log^2\frac{x_{f}}{\epsilon}
- \rho^2\right)sLT.
\ee
Then, the saddle point of the modular parameter $ s $ is given by
\be\label{E:saddle}
	\bar{s} = -i \frac{2}{\rho}\log\frac{x_{f}}{\epsilon}.
\ee
At this saddle point, $ \kappa $ and $ \rho $ are related by $ \kappa = i \rho $ and the
semi-classical partition function of the dyonic magnon is reduced to
\be\label{E:partition}
	e^{iS_{tot}} = \left(\frac{\epsilon}{x_{f}}\right)^{2E},
\ee
where the energy $E$ is mapped to the conformal dimension of the dyonic magnon.
In terms of other variables, $J_1$, $J_2$ and angle difference $ \Delta\phi $ which is
identified with the string worldsheet momentum $p$,
\bea
	J_1 &=& T \int^{L}_{-L}d\sigma\,\sin^2\theta \partial_{\tau}\phi_1
= E - \frac{2T\nu_1}{\sqrt{\nu_1^2-\nu_2^2}} \int^{\pi/2}_{\theta_{min}}d\theta\,
\frac{\sin\theta\cos\theta}{\sqrt{\sin^2\theta- \sin^2\theta_{min}}} ,\nn
	J_2	&=& T \int^{L}_{-L}d\sigma\,\cos^2\theta \partial_{\tau}\phi_2 =
\frac{2T\nu_2}{\sqrt{\nu_1^2-\nu_2^2}} \int^{\pi/2}_{\theta_{min}}d\theta\,
\frac{\sin\theta\cos\theta}{\sqrt{\sin^2\theta- \sin^2\theta_{min}}} ,\nn
	\vert\Delta\phi\vert &\equiv& p = -\int d\phi_1 = 2\int^{\pi/2}_{\theta_{min}}d\theta\,
\frac{\sin\theta_{min}\cos\theta}{\sin\theta\sqrt{\sin^2\theta - \sin^2\theta_{min}}}\,.
\eea
the conformal dimension of the dyonic magnon can be represented as
\be
	E = J_1 + \sqrt{J_2^2 + 4T^2 \sin^2 \frac{p}{2}} .
\ee
This is the exactly known dispersion relation for the dyonic magnon. Especially, when we
set $J_2=1$, the magnon's anomalous dimension obtained from the spin chain model is reproduced.
Moreover, if we choose $\n_2=0$, as mentioned previously, we obtain
$J_2 =0$ and reproduce the magnon's dispersion relation moving on $AdS_5 \times S^2$.

\subsection{Three-point correlation function}

Now, we consider the three-point correlation function between two dyonic magnon
operators ${\cal O}_{m}$ and one marginal scalar operator  $\mathcal{D}_{\chi}$.
Following the AdS/CFT correspondence, the marginal scalar operator is dual to the massless
scalar field in the $AdS$ bulk.  If we turn on the massless scalar field fluctuation, especially
the massless dilaton field fluctuation, in the AdS bulk, then the interaction between the dyonic
string and the massless dilaton field can be described by the Polyakov
action $S_p$
\be
S_p[X,s,\chi] = - \frac{T}{2} \int d^2 \s \ \sqrt{- \g} \g^{\a\b} \pa_{\a} X^A
\pa_{\b} X^B G_{AB} \ e^{\chi/2}  ,
\ee
where $\g_{\a\b}$ is the worldsheet metric and $X=\{z,\vec{x}\}$ represents the coordinates of $AdS$.
Following Ref. \cite{Costa:2010rz}, the three-point correlation function at the saddle point
$\bar{s}$ is given by \cite{Costa:2010rz}
\be
	\langle\mathcal{O}_{m}(0)\mathcal{O}_{m}(x_{f})\mathcal{D}_{\chi}(y)\rangle \approx
\frac{I_{\chi}[\bar{X},\bar{s};y]}{\vert x_{f}\vert^{2E}} ,
\ee
where the subscript `m' means the dyonic magnon.
Here, $I_{\chi} [X,s;y]$ is defined as
\be
	I_{\chi}[X,s;y]= i \int^{s/2}_{-s/2}d\tau\int^{L}_{-L}d\sigma\,
\frac{\delta S_{p}[X,s,\chi]}{\delta\chi}\vert_{\chi=0}\,K_{\chi}(X,s;y),
\ee
where $ K_{\chi}(X,s;y) $ is the bulk-to-boundary propagator of a massless dilaton field
$\chi$ in $AdS$ \cite{Freedman:1998tz}
\be
	K_{\chi}(X , s ; y)=\frac{6}{\pi^2}\left(\frac{z}{z^2+(x-y)^2}\right)^4\,.
\ee
Using the action of the dyonic string, $I_{\chi} [X,s;y]$ can be written in terms of $S_{st}$
in \eq{E:stringaction} 
\be
	I_{\chi}[X,s;y] =  \frac{i}{2}S_{st} K_{\chi}(X , s ; y),
\ee
where $ S_{st} $ is the Polyakov action in the absence of the dilaton fluctuation.
At the saddle point, after performing the integration over the target space variable $\th$,
we can finally obtain
\bea
	I_{\chi}[\bar{X},\bar{s};y]	&=&-\frac{T^2}{\pi^2}\frac{\vert\sin\frac{p}{2}\vert^2}{\sqrt{J_2^2
+4T^2\vert\sin\frac{p}{2}\vert^2}} \ \frac{x_{f}^4}{y^4 (x_{f}-y)^4} .
\eea
Therefore, the three-point correlation function becomes
\be
	\langle\mathcal{O}_{m}(0)\mathcal{O}_{m}(x_{f})\mathcal{D}_{\chi}(y)\rangle
= -\frac{T^2}{\pi^2}\frac{\vert\sin\frac{p}{2}\vert^2}{\sqrt{J_2^2
+4T^2\vert\sin\frac{p}{2}\vert^2}} \frac{1}{x_{f}^{2E-4} y^4 (x_{f}-y)^4}.
\ee
Finally, we can read off the coupling
\be	\la{res:magnoncoupling}
	2\pi^2 a_{Dmm}= -\frac{2T^2\vert\sin\frac{p}{2}\vert^2}{\sqrt{J_2^2
+4T^2\vert\sin\frac{p}{2}\vert^2}}.
\ee

In the gauge theory side, from the conformal dimension of the dyonic magnon given by
\be
	\Delta = J_1+ \sqrt{J_2^2 + 16g^2\vert\sin\frac{p}{2}\vert^2},
\ee
we can evaluate the coupling between two dyonic magnon operators and one marginal scalar
operator by the RG analysis \cite{Costa:2010rz}
\be
	2\pi^2 a_{Dmm} = -g^2\frac{\partial}{\partial g^2}\Delta = - \frac{8g^2\vert\sin\frac{p}{2}\vert^2}{\sqrt{J_2^2 + 16g^2\vert\sin\frac{p}{2}\vert^2}}.
\ee
Using \eq{rel:tension}, this result is exactly the same as the string calculation
\eq{res:magnoncoupling}
in the semi-classical limit. If we set $J_2=1$, this result is reduced to the coupling
of the magnon in the spin chain model. Furthermore, we set $J_2=0$ the above becomes
the string moving in $AdS_5 \times S^5$, which is equivalent to the coupling
between two magnon operators and one marginal scalar operator at the large t' Hooft coupling limit.

\section{Dyonic spike}

For the dyonic spike, the energy and the angle difference usually have infinite value while
two angular momenta $J_1$ and $J_2$ are finite. In order to describe this dyonic spike solution,
we should impose appropriate boundary conditions.
We first impose $c_2 =0$ for finiteness of the second angular momentum and assume that there
exists a maximum value $\th_{max}$ with
$\th_{max}'=0$.
Second, in order to make the first angular momentum finite we impose $\pa_{\ta} \ph_1 = 0$
at $\th_{max}$, which was also used in investigating the spike moving on $AdS_5 \times S^2$
\cite{Park:2010vs}. The last boundary condition, $\pa_{\ta} \ph_1 |_{\th_{max}}= 0$,
determines the maximum value of $\th$ in terms of other parameters
\be
\sin^2\theta_{max} = \frac{ac_1}{b^2 \nu_1}.
\ee
From \eq{E:htheta} together with $\th_{max}$, the first boundary condition, $\th_{max}'=0$,
is satisfied as the integration constant $w^2$ is given by
\be
w^2 = \frac{ac_1}{\nu_1}(\nu_1^2-\nu_2^2) + \frac{b^2\nu_1c_1}{a} .
\ee
Substituting this integration constant into \eq{E:htheta}, the equation for $\th$ can
be rewritten as
\be
\theta'^2 = \frac{b^2(\nu_1^2-\nu_2^2)}{(a^2-b^2)^2\sin^2\theta}\left(\sin^2\theta_{max}
-\sin^2\theta\right)\left( \sin^2\theta - \sin^2\theta_{min}\right),		
\ee
where $\th_{min}$ is given by
\be
\sin^2\theta_{min} =  \frac{b^2\nu_1^2}{a^2(\nu_1^2-\nu_2^2)} .
\ee
Note that because $0 < \sin^2\theta_{min} < 1$, at least $\n_1 > \n_2$ and $a > b$ when
we assume that all parameters are positive.

\subsection{Two-point correlation function}
Following the same procedure in the previous section, we first calculate convoluted action
for the dyonic spike $\bar{S}$, which is
\be
	\bar{S} = -\frac{T}{2}\int d^2\sigma\, \frac{b^2}{a^2}\nu_1^2 \equiv
-\frac{T}{2}\int d^2\sigma\, \rho^2.
\ee
Then, we can reobtain the same expressions for the saddle point \eq{E:saddle}
and the semi-classical partition function \eq{E:partition}. For dyonic spike,
the various conserved charges, the energy, angular momenta and angle difference, are given by
\bea
	E &=& T\int^{L}_{-L}d\sigma\, \rho = \frac{2T\nu_1(a^2-b^2)}{ab\sqrt{\nu_1^2-\nu_2^2}}
\int^{\pi/2}_{\theta_{min}} d\theta\, \frac{\sin\theta}{\cos\theta}\frac{1}{\sqrt{\sin\theta
-\sin^2\theta_{min}}} , \nn
	J_1 &=& T\int^{L}_{-L}d\sigma\, \sin^2\theta \partial_{\tau}\phi_1
= -\frac{2T\nu_1}{\sqrt{\nu_1^2-\nu_2^2}}\int^{\pi/2}_{\theta_{min}}d\theta
\frac{\sin\theta\cos\theta}{\sqrt{\sin\theta-\sin\theta_{min}}} ,\nn
	J_2 &=& T\int^{L}_{-L}d\sigma\, \cos^2\theta \partial_{\tau}\phi_2
= -\frac{2T\nu_2}{\sqrt{\nu_1^2-\nu_2^2}}\int^{\pi/2}_{\theta_{min}}d\theta
\frac{\sin\theta\cos\theta}{\sqrt{\sin\theta-\sin\theta_{min}}} ,\nn
	\vert\Delta\phi\vert &\equiv& p = -\int d\phi_1= \frac{E}{T}- 2
\int^{\pi/2}_{\theta_{min}}d\theta\,\frac{\cos\theta\sin\theta_{min}}{\sin\theta
\sqrt{\sin^2\theta-\sin^2\theta_{min}}}.
\eea
From these definitions of the conserved charges, we can rewrite the dyonic spike's energy
in terms of other quantities, which gives the dispersion relation of the dyonic spike
\be
	E = T \vert\Delta\phi\vert + 2T \tilde{\theta} ,
\ee
with
\be
	\tilde{\theta}\equiv \frac{\pi}{2} - \theta_{min} = \arcsin\frac{\sqrt{J_1^2-J_2^2}}{2T} .
\ee
If we set $\n_2 = 0$, then $J_2$ vanishes and the above dispersion relation is reduced to
one for the spike moving on $AdS_5 \times S^2$.

\subsection{Three-point correlation function}
Let us now consider the three-point correlation function between two single dyonic spike
operators $ \mathcal{O}_{s} $ and one marginal operator $ \mathcal{D}_{\chi} $ which is dual
to a massless dilaton field $\chi$. We first insert the dyonic spike solution into the
string action and perform the integration at the saddle point where $ \kappa = i\rho $ is valid,
\bea
	S_{st} &=& \frac{T}{2}\int^{\bar{s}/2}_{-\bar{s}/2}d\tau \int^{L}_{-L}d\sigma \,
 2b^2\left(\frac{\nu_1^2-\nu_2^2}{a^2-b^2}\cos^2\theta - \frac{\nu_1^2}{a^2}\right) \nn
 		   &=& \int^{\bar{s}/2}_{-\bar{s}/2}d\tau\, \ls -\rho E + 2T\sqrt{\nu_1^2-\nu_2^2}
 \cos\theta_{min} \rs
\eea
Then, we can obtain
\bea
I_{\chi}[\bar{X},\bar{s};y] &=& \frac{3i}{\pi^2}S_{st} K_{\chi}(X , s ; y) \nn
								&=& \frac{1}{2 \pi^2} \ls -\frac{1}{2} T\vert\Delta\phi\vert
- T\tilde{\theta}+ T\tan\tilde{\theta}_{min} \rs \frac{x^4_{f}}{y^4 (x_f-y)^4},
\eea
and the three-point correlation function between two dyonic spike operators $\mathcal{O}_{s}$
and a marginal scalar operator becomes
\be
\langle\mathcal{O}_{s}(0)\mathcal{O}_{s}(x_{f})\mathcal{D}_{\chi}(y)\rangle
		= \frac{1}{2\pi^2} \ls -\frac{1}{2}T\vert\Delta\phi\vert - T\tilde{\theta}
+ T\tan\tilde{\theta}_{min} \rs \frac{1}{x^{2E-4}_{f} y^4 (x_f-y)^4}\,.
\ee
Finally, we read off the coupling
\be\label{E:sadds}
	2\pi^2 a_{Dss} = -\frac{1}{2}T\vert\Delta\phi\vert - T\tilde{\theta}+ T\tan\tilde{\theta}_{min}\,.
\ee
Using the RG analysis, we can also calculate the coupling in the gauge theory side
\bea
 2\pi^2 a_{Dmm} &=& -g^2\frac{\partial}{\partial g^2}E =- \frac{T}{2}\frac{\partial}{\partial T}
                    \left(T\vert\Delta \phi\vert + 2T \arcsin\frac{\sqrt{J_1^2-J_2^2}}{2T}\right)\nn
                &=& -\frac{1}{2}T\vert\Delta\phi\vert - T\tilde{\theta}+ T\tan\tilde{\theta}_{min},
\eea
which is exact same as \eq{E:sadds} obtained from semi-classical string calculation.
The above can be also reduced to the spike moving on $AdS_5 \times S^2$ in the limit of $J_2=0$.

\section{Discussion}

In this paper, we investigated the dyonic string in the infinite size limit moving in
$AdS_5 \times S^3$, which has three conserved quantities, $E$, $J_1$ and $J_2$.
In the parameter region $a < b$, if imposing that $J_2$ is finite, the dyonic string
is dual to the dyonic magnon of the gauge theory which describes the bound state of magnons.
Following the Janik's proposition, we reproduced the known dispersion relation of the
dyonic magnon. In addition, we calculated the three-point correlation function between two
dyonic magnons and one marginal operator and showed that the coupling of them in the string
set-up is exactly the same as one obtained by the RG analysis in the gauge theory side.
Furthermore, if we set $\n_2 = 0$, $J_2$ becomes zero and the results of the dyonic string
is reduced to ones of the string moving in $AdS_5 \times S^2$. Especially, if we set $J_2 =1$,
the results of the dyonic string can exactly describe ones of the magnon in the spin chain
model in the arbitrary t' Hooft coupling.

We also investigated the two- and three-point correlation functions of the dyonic string
for $a>b$, in which the dual object is the dyonic spike. After some calculation, we
reproduced the known dispersion relation of the dyonic spike and calculated the coupling
between two spikes and one marginal operator. As we expected, the coupling in the string
calculation coincide with the result of the gauge theory.

\vspace{1cm}

{\bf Acknowledgement}

This work was supported by the National Research Foundation of Korea(NRF) grant funded by
the Korea government(MEST) through the Center for Quantum Spacetime(CQUeST) of Sogang
University with grant number 2005-0049409. C. Park was also
supported by Basic Science Research Program through the
National Research Foundation of Korea(NRF) funded by the Ministry of
Education, Science and Technology(2010-0022369).

\end{document}